# The variable quality of metadata about biological samples used in biomedical experiments

Rafael S. Gonçalves and Mark A. Musen

Stanford Center for Biomedical Informatics Research
Stanford University, CA, USA

Corresponding author: Rafael S. Gonçalves (rafael.goncalves@stanford.edu)

**Abstract**
We present an analytical study of the quality of metadata about samples used in biomedical experiments. The metadata under analysis are stored in two well-known databases: BioSample—a repository managed by the National Center for Biotechnology Information (NCBI), and BioSamples—a repository managed by the European Bioinformatics Institute (EBI). We tested whether 11.4M sample metadata records in the two repositories are populated with values that fulfill the stated requirements for such values. Our study revealed multiple anomalies in the metadata. Most metadata field names and their values are not standardized or controlled. Even simple binary or numeric fields are often populated with inadequate values of different data types. By clustering metadata field names, we discovered there are often many distinct ways to represent the same aspect of a sample. Overall, the metadata we analyzed reveal that there is a lack of principled mechanisms to enforce and validate metadata requirements. The significant aberrancies that we found in the metadata are likely to impede search and secondary use of the associated datasets.

## Introduction

The metadata about scientific experiments are essential for finding, retrieving, and reusing the scientific data stored in online repositories. Finding relevant scientific data requires not only that the data simply be accompanied by metadata, but also that the metadata be of sufficient quality for the corresponding datasets to be discovered and reused. When the quality of the metadata is poor, software systems that index and avail themselves to the experimental data may not find and return search results that otherwise would be appropriate for given search criteria. In addition, significant metadata post-processing efforts may be required to facilitate data analysis.

The literature on metadata quality generally point to the need for better practices and infrastructure for authoring metadata. Bruce *et al.*[1] define various metadata quality metrics, such as completeness (e.g., all necessary fields should be filled in), accuracy (e.g., the values filled in should be specified as appropriate for the field), and provenance (e.g., information about the metadata author). Park *et al.*[2,3] specified several high-level principles for the creation of good-quality metadata. The metrics mentioned in these works have been recently supplemented by the FAIR data principles[4]. The FAIR principles specify desirable criteria that metadata and their corresponding datasets should meet to be Findable, Accessible, Interoperable, and Reusable.

Several empirical studies suggest that metadata quality needs to be significantly improved. Infrequent use of ontologies to control field names and values and lack of validation have been identified as key problems. For example, Zaveri *et al*. reported that metadata records in the Gene Expression Omnibus (GEO) suffer from redundancy, inconsistency, and incompleteness[5]. This problem occurs because GEO allows users to create arbitrary fields that are not predefined by the GEO data dictionary and it also does not validate the values of those fields. Hu *et al.* developed an agglomerative clustering algorithm to clean metadata field names—cutCluster[6]. Hu *et al.* tested whether a sample of 359 out of 11,000 field names in GEO metadata were clustered similarly to a gold standard clustering crafted by the authors. Hu *et al.* concluded that multiple field names would require human verification to determine their correct cluster. Park[7] examined the use of Dublin Core (DC) elements in field names, in a corpus of 659 metadata records sampled from digital image collections on the Web. Park identified various problems with the representation of DC elements that could have been prevented with better infrastructure to map metadata field names to DC elements. Bui *et al.*[8] conducted a similar study to investigate the use of DC elements in metadata fields in a larger corpus of around 1 million records. The authors found that 6 DC fields are "rather well populated," while the other 10 fields that they analyzed were poorly populated. However, none of these authors investigated whether the content of metadata fields is appropriately specified according to the fields' expected values. For example, the studies checked whether DC fields are populated but not whether the values for the *dc:date* field are dates formatted according to some standard, or whether the values for the language field resolve to controlled terms in an ontology about languages or a language value set. For data to be FAIR, the value of each metadata field needs to be accurate and uniform (e.g., relying on controlled terms where possible), and to adhere to the field specification. Using controlled terms as a means to standardize metadata field names and field values allows users to be able to find data in a principled way, without having to cater to *ad hoc* representation mechanisms.

In this paper, we present an analysis of the quality of metadata in two online databases: the NCBI BioSample[9], which is maintained by the U.S. National Center for Biotechnology Information (NCBI), and the EBI BioSamples[10,11], which is maintained by the European Bioinformatics Institute (EBI). These databases store metadata that describe the biological materials (samples) under investigation in a wide range of projects. We selected the NCBI BioSample as it represents the most recent NCBI metadata repository initiative, and because the NCBI BioSample repository was designed to standardize sample descriptions across all NCBI repositories (including GEO) with a focus on the use of controlled terms from ontologies. The EBI BioSamples was selected because it is EBI's equivalent repository to the NCBI BioSample, and because the EBI curates the metadata in its repository, contrary to the NCBI. The curation process used with the EBI BioSamples repository involves mapping values to controlled terms, and it results in simpler metadata (e.g., values such as "N/A" or "missing" are pruned) that are presumably closely aligned with ontologies. The metadata in the EBI BioSamples are partially-curated descriptions of sample-related data hosted in databases such as ArrayExpress[12]. ArrayExpress includes all the microarray data in the GEO database. Thus, in our study, we expect to find that there is overlap in the contents of the EBI BioSamples and the NCBI BioSample repositories, and that the curated EBI BioSamples metadata are of higher quality.

## Methods

Our goal is to measure the quality of metadata records based on whether the fields that the records describe comply with their specification. We consider metadata to be of good quality if the metadata fields use controlled terms when indicated, if their values are parseable, and if the values match the expectations of the database designers. We analyzed metadata fields (so-called *attributes*) that have computationally verifiable expectations for their values in the two repositories for metadata about biomedical samples. For example, a field that is expected to be populated with numeric values can be unambiguously verified, while a field that is populated with free-text values would pose a non-trivial challenge for automated verification. A metadata attribute comprises a pair consisting of an attribute name and an attribute value. For example, values for the attribute named *disease* of human samples in the NCBI BioSample should correspond to terms in the Human Disease Ontology (DOID), according to BioSample documentation. We used the BioPortal repository of publicly available biomedical ontologies[13] to identify correspondences between metadata values and ontology terms.

We acquired a copy of the NCBI BioSample database from the central NCBI FTP archive[14] on June 25, 2017. The BioSample database was distributed as an XML file, with no explicit versioning information. Our copy of NCBI's BioSample contained 6,615,347 metadata records. A typical BioSample record appears in **Figure 1**.

The EBI software infrastructure did not have a downloadable archive containing the entire BioSamples database. We obtained a snapshot of the database on November 15, 2017 by contacting the EBI IT Helpdesk. Our copy of EBI's BioSamples contained 4,793,915 metadata records. In this paper, when we refer to the metadata in either NCBI BioSample of EBI BioSamples, our comments are necessarily based on the snapshots that we obtained of the two repositories in 2017.

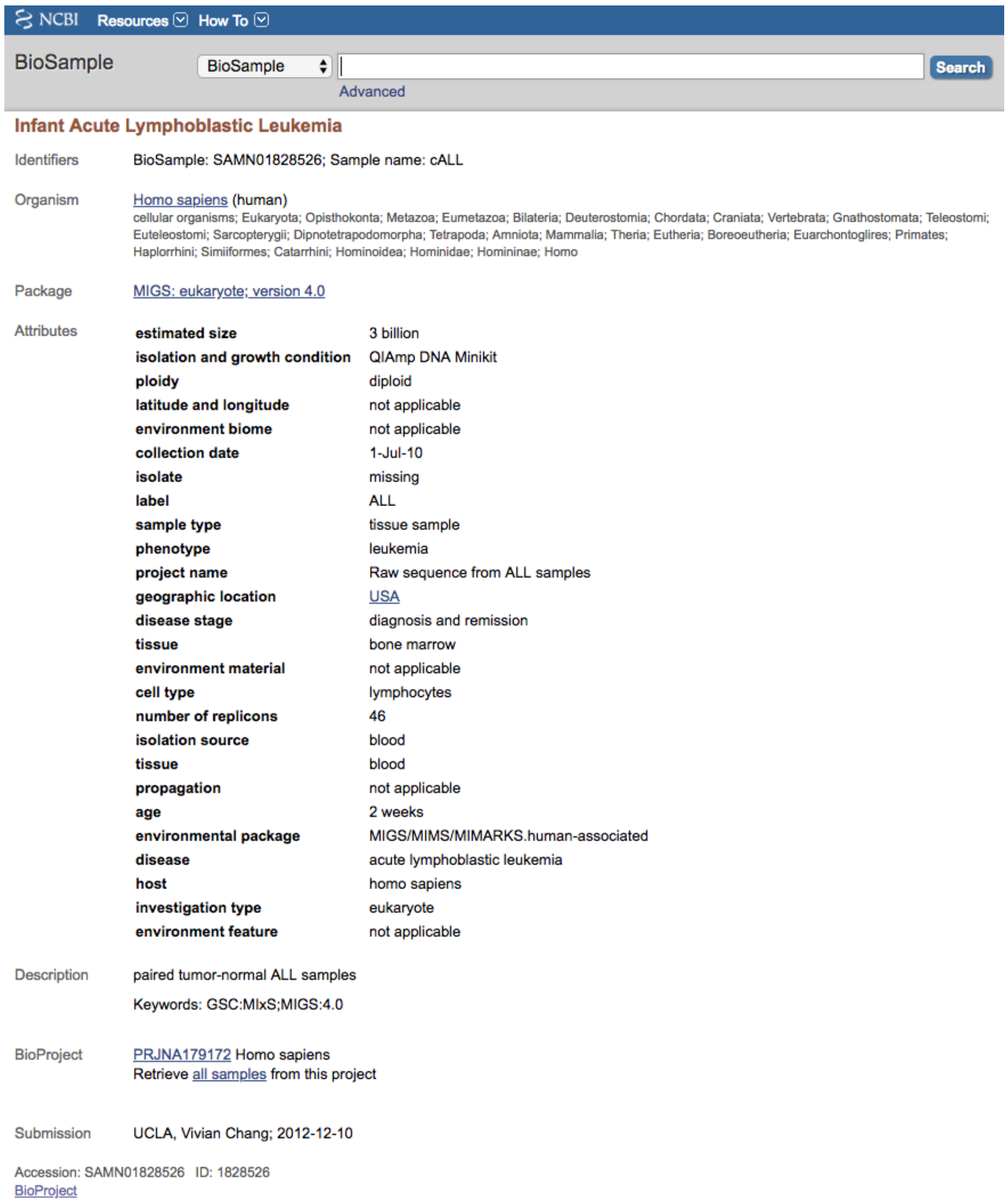

**Figure 1. Example metadata record from the NCBI BioSample.** An NCBI BioSample metadata record has a title, potentially multiple identifiers associated with it, an organism, a package specification (explained in Section 2.1), multiple attributes in the form of name-value pairs, a description with keywords associated with it, information about the record submitter, and finally accession details.

We built a software tool to extract key bits of information about each metadata record in the samples databases and to determine whether the attributes of each sample record were filled in and well-specified. Our tool collects the following data: sample identifier, accession number, publication date, last update date, submission date, identifier and name of the sample organism, owner name, and package name. Then, for each attribute within our tested attributes, the software records the attribute name, its value, and verifies whether it is filled in according to the attribute's specification. An attribute specification describes the format and content of the expected attribute value. Each repository defines and documents its own attribute specifications. We developed our software to determine whether these specifications hold in the metadata that we processed.

We built a second tool to cluster a list of given strings according to their similarity using the affinity propagation clustering algorithm[15]. Affinity propagation is a machine learning algorithm that identifies exemplars among data points and creates clusters of data points around the exemplars. This clustering technique is desirable for our study because it does not require specifying the number of clusters upfront (which are unknown in our case), and because it computes a representative value for each cluster (the *exemplar*). We used the implementation of the affinity propagation algorithm in the scikit-learn Python package[16]. To compute the similarity between strings we use the Levenshtein edit distance. The Levenshtein distance between two strings $s$ and $t$ is the shortest sequence of single-character edit commands (insertions, deletions, or substitutions) that transforms $s$ into $t$. We chose this distance metric because it is widely used in spell-checkers and search systems, it accounts for simple typing errors, and it is not restricted to strings of equal length.

**Code availability**
All code used for the quality assessment of metadata and the clustering of metadata keys is available at https://github.com/metadatacenter/metadata-analysis-tools.

**Data availability**
The data used and generated throughout the study described in this paper are available in Figshare[17] at https://doi.org/10.6084/m9.figshare.6890603.

## NCBI BioSample Overview

Officially launched in 2011, the NCBI BioSample repository accepts submissions of metadata through a Web-based portal that guides users through a series of metadata-entry forms. The first form prompts users to choose a *package*. A package represents a type of sample and it specifies a set of attributes that should be used to describe samples of a particular type. For instance, the *Human.1.0* package requires its records to have the attributes *age*, *sex*, *tissue*, *biomaterial provider*, and *isolate*. This package also lists other attributes that can be optionally provided. Each of the 104 BioSample package types has a different set of rules regarding which attributes are required and which are optional[18]. A notable exception is the *Generic* package, which has no requirements at all. This package is not listed in the online package documentation and it is not an option in BioSample's Web forms.

A metadata record defines multiple attributes, each of them composed of an attribute name and a value. BioSample provides a dictionary of 452 metadata attribute names[19] that can be used to describe the samples that form the substrates of experiments. Metadata authors can, however, provide additional attributes with arbitrary names with no guidance or control from BioSample. Each metadata record describing a sample can contain multiple attributes. Given the biomedical domain, we expect BioSample metadata to use terms from ontologies in BioPortal—a repository that currently hosts over 700 publicly available biomedical ontologies.

**Analysis of NCBI BioSample metadata**

Our study assesses the quality of metadata in BioSample according to whether the attributes in the metadata records specify (1) a controlled attribute name (i.e., provided by an ontology or other controlled term source), (2) an attribute name that is in BioSample's attribute dictionary, and (3) a valid value according to the attribute specification. We analyzed all the BioSample attributes and categorized those attributes that have the same type of expected values into the groups described in the following subsections.

**Ontology-term attributes**. There are 9 BioSample attributes that dictate the use of term values from specific ontologies. For example, the attribute *phenotype*, representing the phenotype of the sampled organism, should have input values that are terms from the Phenotypic Quality Ontology (PATO), according to the BioSample documentation. To verify whether ontology terms supply values for attributes in BioSample when appropriate, we performed searches in BioPortal for exact matches of the possible values for each relevant BioSample attribute field within the ontology that the BioSample attribute documentation indicates should provide values for that field. We indicate that an ontology-term attribute is *well-specified* if its value matches a term in the designated ontology. When matching terms, the algorithm implemented in BioPortal takes into consideration the term names, synonyms, and term identifiers.

**Value set attributes.** There are 32 attributes whose values are constrained to value sets specified in the BioSample documentation. For example, the attribute name *dominant hand* takes on values from a value set composed of the terms *left*, *right*, and *ambidextrous*. We developed methods for verifying that values stored for each of these types of attributes are appropriate, that is, whether values match against terms in the corresponding value sets. We tested whether the values found in BioSample records actually corresponded to the values defined in value sets in the BioSample documentation.

**Boolean attributes.** We tested 4 attributes in BioSample packages that require a Boolean value. We indicate that a Boolean attribute is *well-specified* if its value is *true* or *false*, regardless of capitalization. We consider values such as *f* or *yes* to be invalid.

**Integer attributes.** We tested 4 attributes that require an integer value. An integer attribute is *well-specified* if the given value can be parsed as an integer.

**Timestamp attributes.** We tested 11 attributes that require a timestamp value. A timestamp attribute is *well-specified* if the given value is in the format "DD-Mmm-YYYY", "Mmm-YYYY" or "YYYY" (e.g., 20-Nov-2000, Nov-2000 or 2000), or adheres to the ISO 8601 standard for timestamps: "YYYY-mm-dd", "YYYY-mm" or "YYYY-mm-ddThh:mm:ss" (e.g., 2000-11-20, 2000-11 or 2000-11-20T17:30:20).

We gathered similar information about other structured attributes, although we did not test the validity of those values in the BioSample data. For example, there are 161 attributes that require a unit of measure, 21 attributes that require a PubMed ID, and so on.

We chose to validate the 5 groups above because the characteristics of these groups are easily tested and because the expected values of the attributes are straightforward for users to specify (e.g., compared to attributes such as those that require a value to be composed of a floating-point number followed by a special symbol).

## EBI BioSamples Overview

The EBI BioSamples repository stores metadata about biological samples used in experiments registered, for example, in ArrayExpress. Human curators use a software system known as Zooma[20] to standardize the metadata and to add them to BioSamples. Zooma maps free text annotations to terms in ontologies hosted in the EBI's Ontology Lookup Service (OLS)[21–23]. The tool applies these mappings based on rules that are learned from the manual curation carried out in the ArrayExpress repository.

Metadata authors can submit metadata to EBI BioSamples in the form of SampleTab files. The SampleTab file format is a tab-delimited, spreadsheet-like format composed of two sections: Meta-Submission Information (MSI) and Sample Characteristics Description (SCD). The MSI section contains information about the submission (e.g., title, identifier, description, version), about the submitting organization (e.g., name, address), and about database links (e.g., the name of the database and the identifier within that database). The SCD section describes the sample characteristics via attributes of the form of name–value pairs. For the purposes of our study, we focused on the SCD section. Metadata can be submitted to the EBI BioSamples via Web forms or via REST APIs.

BioSamples specifies only 3 attribute names (so-called "named attributes") that should be used to describe samples[24], whereas NCBI BioSample specifies 452. The named attributes that have a definition in the EBI repository documentation are:

- **Organism** – "Value should be scientific name and have NCBI Taxonomy as a Term Source REF with associated Term Source ID."
- **Sex** – "Prefer 'male' or 'female' over synonyms. May have other values in some cases e.g. yeast mating types."

There is no definition for the expected values of attributes using the **Material** named attribute. In addition to named attributes, BioSamples allows metadata submitters to use "free-form attributes" to describe samples (i.e., attributes containing *ad hoc* attribute names other than the 3 discussed above).

We carried out the same analysis of the EBI BioSamples repository that we performed for NCBI BioSample. In our study of the EBI BioSamples, we assess the quality of the 3 metadata attributes defined in the BioSamples documentation as follows:

- An **Organism** attribute is well-specified if the value corresponds to a term in the NCBI Taxonomy.
- A **Material** attribute is well-specified if the value corresponds to a term in a biomedical ontology.
- A **Sex** attribute is well-specified if the value is in the NCBI value set for the *sex* attribute, which includes the EBI-preferred terms "male" and "female".

In the case of the **Sex** attribute, since there is no pre-defined range for the values in EBI BioSamples, we used the value set defined in the NCBI BioSample documentation. We used this value set to be able to compare results between the two repositories.

## Results

We analyzed whether the values of metadata attributes comply with the specifications set out by the developers of each of the two hosting databases, NCBI BioSample and EBI BioSamples. We evaluated the quality of metadata records that exist in both databases, which we determined according to their accession identifiers. Finally, we clustered the metadata attribute names to identify redundant attribute names used to represent the same aspect of a sample, and thus ideally could be denoted by a single attribute name.

### NCBI BioSample

The metadata records in BioSample represent 94 unique package types. Thus, not all of the 104 BioSample packages types are used. Generic packages make up the bulk of the BioSample database—85% of the records use this package definition (**Figure 2**). The next most populated package is *Pathogen*, consisting of 3.2% of the records.

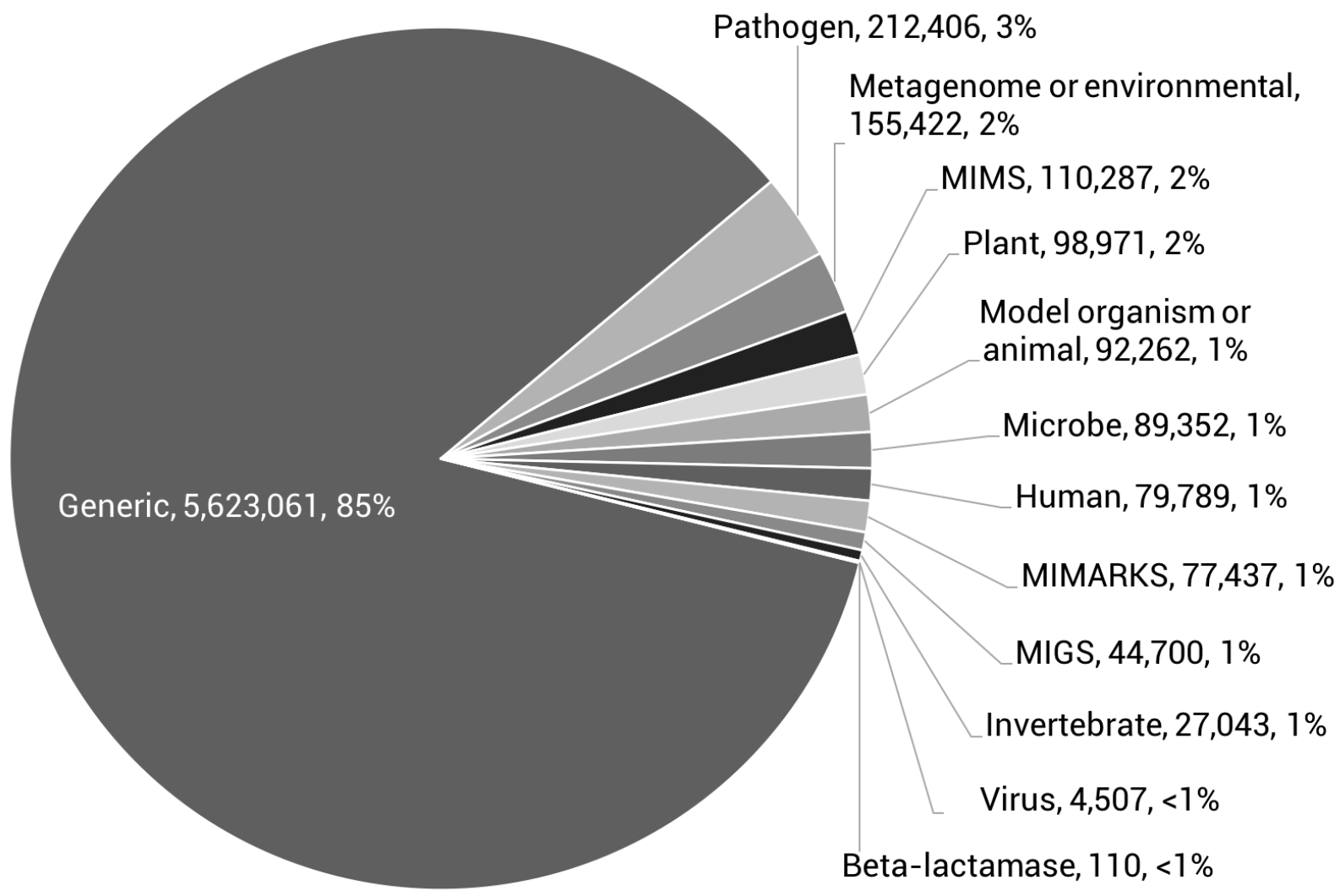

**Figure 2. Mention of metadata packages in NCBI BioSample.** The chart shows the package names followed by the number (and percentage) of metadata records that use that package. The *Generic* package does not specify any required or optional attributes.

We examined the evolution of the number of Generic versus non-Generic submissions to BioSample over the years, to determine whether the metadata records adhering to the Generic package were legacy submissions potentially imported from other databases. In **Figure 3**, we show the total number of metadata record submissions to NCBI BioSample from 2009 to 2017. Nearly all of the submissions until 2013 used the Generic package. After 2013, one observes some adoption of packages other than the Generic one, although most metadata (between 75% and 80% of all records) were still submitted using the Generic package between 2014 and 2017.

BioSample records contain a total of 82,360,966 attributes (name–value pairs). Attribute names either are selected from the BioSample dictionary or are user-defined. A total of 12,284,229 pairs (15% of the total), encompassing over 2,303,021 metadata records (35% of all records), use attribute names that are not specified in the BioSample attribute dictionary. Of these attributes, we identified 18,198 syntactically unique custom attribute names specified by submitters. For example, some records use the name *Altitude (m)* instead of *altitude*—the attribute name defined by BioSample. The records that contain these attributes have been submitted by 313 different laboratories. Overall, there are 18,650 different attribute names used in BioSample metadata records—452 are BioSample-specified (2.4%), and the remainder are user-specified (97.6%). Only 9 of the 452 BioSample-specified attribute names are terms that are taken from standard ontologies. It is unclear whether any of the user-defined attribute names corresponds to ontology terms; in our analysis, we did not find any values for user-defined attribute names that correspond to ontology term IRIs (either in their full or prefixed form, e.g., *ENVO:00000428*). Of all BioSample records, only 197,123 Generic-package records (0.03%) do not specify any attributes. On average, each BioSample metadata record specifies 12 attributes. The vast majority of BioSample records (97%) specify at least one attribute.

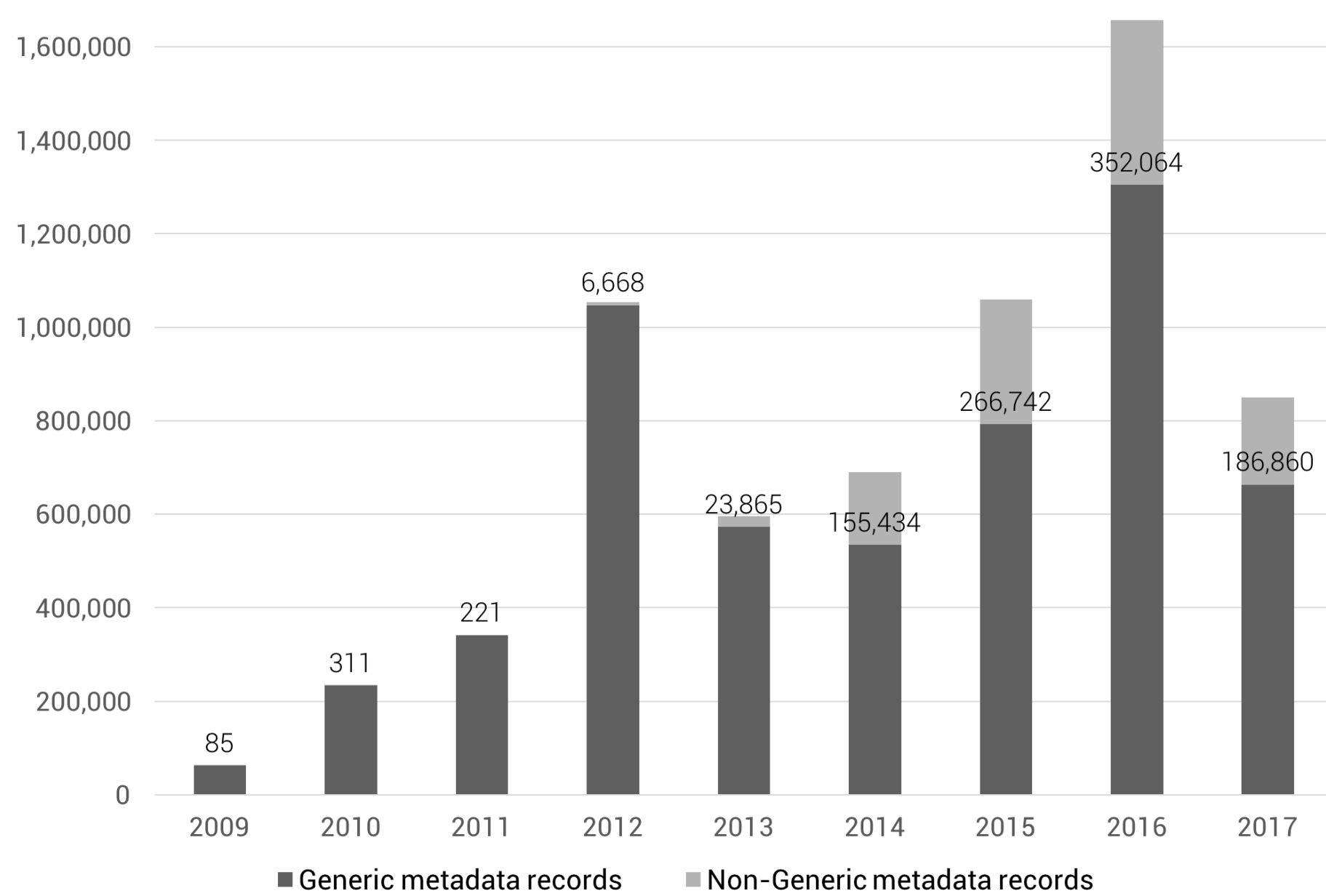

**Figure 3. Metadata submissions to NCBI BioSample from 2009-2017.** The columns represent the total number of metadata record submissions to NCBI BioSample in a year, split between Generic and non-Generic records. The *Non-Generic metadata records* column contains data labels with the absolute number of records. Generic records make up nearly all the submissions in the early years of BioSample, and the bulk of the submissions even in recent years.

The primary results of our study of NCBI BioSample metadata are presented in **Figure 4**. We will now explain each of the columns in the Figure in the order of their appearance.

**Ontology-term attributes**. Most attributes in BioSample whose values are intended to be taken from terms in standard ontologies do not contain terms from ontologies. There are 1,016,483 records (15.4%) that contain a value for one or more attributes that ideally require an ontology term. Out of those, only 441,719 (43% of this subset) have valid values for their ontology-term attributes. These records contain a total of 1,976,642 ontology-term attributes, and only 639,154 (32%) of those attributes contain values that are actually ontology terms. Some values for these attributes do not match with terms in BioPortal because they are not typed correctly or contain non-alphabetic symbols. For example, the *disease* attribute requires a term from the Human Disease Ontology (DOID), but some values given include *gastrointestinal stromal tumor_4* (*gastrointestinal stromal tumor* is a class in DOID), *HIV_Positive* (*HIV* is a class in DOID), *infected with Tomato spotted wilt virus isolate p105RBMar*, which does not have a close match, *lung_squamous_carcinoma,* which would have matched with a term if not for the underscores, numeric values that do not match BioPortal terms, and so on.

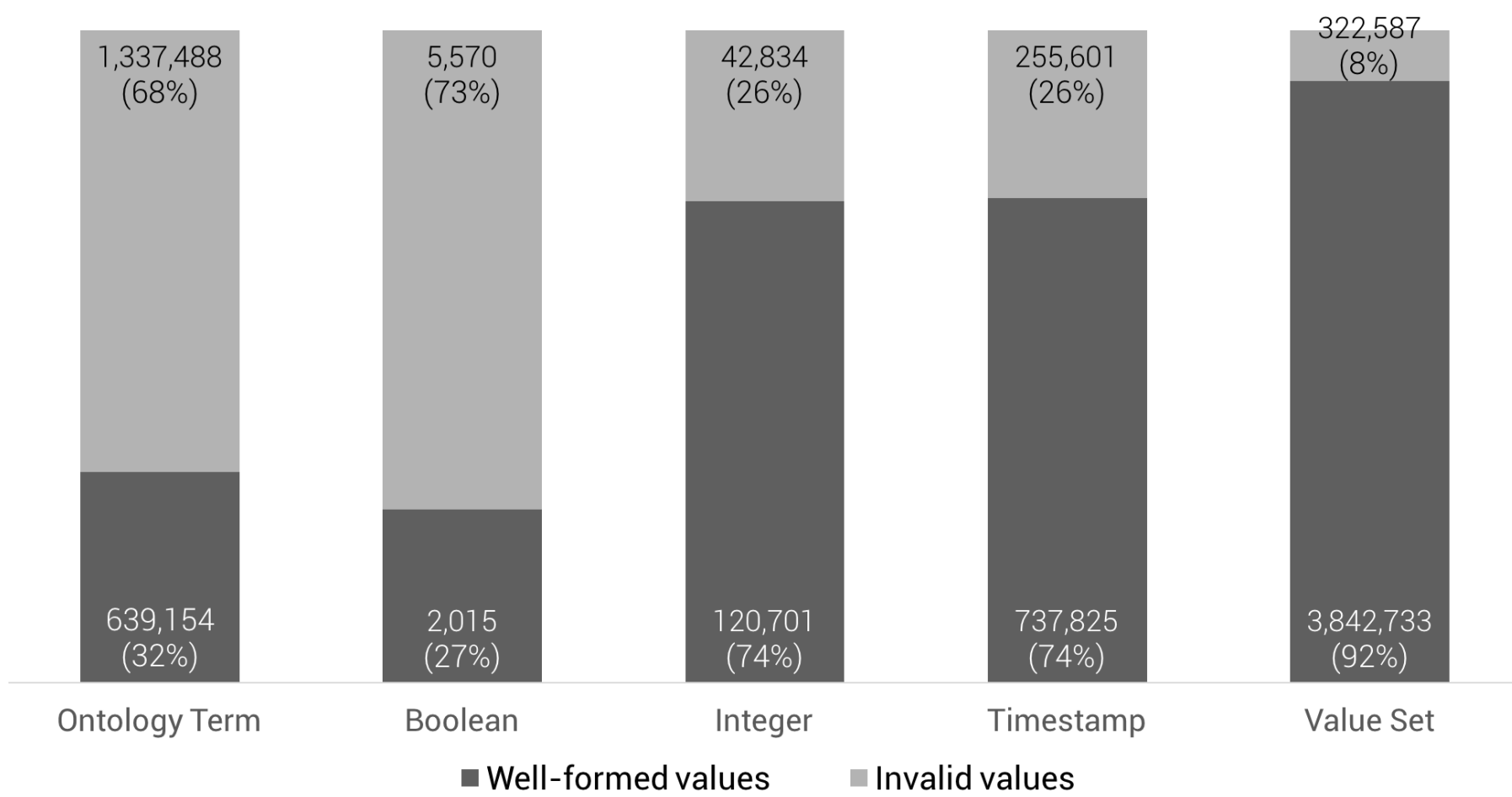

**Figure 4. Quality of dictionary attributes in NCBI BioSample according to their type**. The columns show the number and percentage of attributes whose values are well-specified or invalid.

**Value-set attributes**. Among the attribute groups we analyzed, the attributes that use value sets are the most well-specified. There are 4,028,758 records that contain one or more attributes whose values are intended to be taken from value sets. Of those, 3,781,283 records (94%) contain values that appear to be valid. These records specify a total of 4,165,320 value-set attributes, and 3,842,733 (92%) of those are well-specified. Even though most records adhere to the value sets, we observed that a wide range of values is given for even seemingly straightforward attributes such as *sex*. This attribute has possible values *male*, *female*, *pooled male and female*, *neuter*, *hermaphrodite*, *intersex*, *not determined*, *missing*, *not applicable*, and *not collected*. The values in BioSample records include invalid variations of the accepted values *male* and *female,* such as *m* and *f*, as well as acceptable variations, such as *Male* and *FEMALE* (our matching algorithm ignores letter case). Other values include *pool of 10 animals; random age and gender*, *juvenile*, *Sexual equality*, *parthenogenic*, *larvae, pupae and adult (queens - workers)*, *castrated horse*, *gynoparae*, *uncertainty*, *Vaccine and Infectious Disease Division*, *Clones arrayed from a variety of cDNA libraries*, and *Department I of Internal Medicine*, values containing only numbers, only symbols, and misspelled words such as *mal e*, *makle*, and *femLE*.

**Boolean attributes**. The Boolean attributes are the most inconsistent of all the attributes we analyzed. Overall, 6,767 BioSample records contain a value for one or more Boolean attributes. Only 2,013 (30%) of those records have attributes that are valid. These records specify 7,585 Boolean-type attributes, of which only 2,015 (27%) are well-specified. For example, for the *smoker* attribute, there are such diverse values as: *Non-smoker*, *nonsmoker*, *non smoker*, *ex-smoker*, *Ex smoker*, *smoker*, *Yes*, *No*, *former-smoker*, *Former*, *current smoker*, *Y*, *N*, *0*, --, *never*, *never smoker,* among others.

**Integer attributes**. The values for integer attributes are mostly well-specified. There are 158,854 records containing one or more attributes that require an integer value. Out of those, 120,026 records (76%) contain valid attributes. These records specify a total of 163,535 integer attributes, and 120,701 of those (74%) are well-specified. The NCBI-specified attribute *medication code,* which is intended to be an integer, does not have any valid values in the repository (values include *Insulin glargine injectable solution, Insulin lispro injectable solution, Fluoxetine, Simvastatin, Isosorbide mononitrate, Amlodipine/Omelsartan medoxomil*). The BioSample documentation does not specify why the *medication code* attribute should take numeric values. The attribute *host taxonomy ID*, which should be filled in with integers corresponding to entries in the NCBI taxonomy, has values such as *e;N/A*, *Mus musculus,* and *NO*. The expected integer values for the *host taxonomy ID* attribute are identifiers that should correspond to organism names in the NCBI organismal classification.

**Timestamp attributes**. The timestamp attributes are generally well-specified. There are 2,913,038 metadata records containing one or more attributes that require a timestamp value. These records specify nearly 1 million metadata attributes, out of which there are 737,825 (74%) whose values match one of the expected date formats or the ISO standard. Among the invalid values we found wrongly formatted dates such as: *1800/2014*, or *Jan-Feb 2009*, and text values such as *no description*, or *unspecified*.

We found that the quality of the metadata attribute values in records that adhere to packages other than the Generic package is actually inferior to that of the overall metadata quality. In **Figure 5** we show the quality of the metadata in the packaged subset of metadata records. The quality of the attributes is inferior in all attribute groups except the *Timestamp* group (and only by 1%). The Boolean, value-set, and integer groups of attributes are significantly inferior in quality compared to the overall quality across the repository. Our expectation was that submitters who put in the effort to select and adhere to a specific metadata package would likely produce higher-quality metadata. This turned out to be false.

## Analysis of metadata in both EBI and NCBI repositories

We compared the sets of metadata record identifiers in the NCBI BioSample and the EBI BioSamples, to discover there are 2,913,038 records that exist in both databases. This is because EBI BioSamples consumes metadata from ArrayExpress, which contains metadata from GEO, and because GEO metadata are contained in the NCBI BioSample. A large proportion of these common records specify "EBI" as the value for the *Owner* metadata field (1,220,429 records, 42%). Metadata records with the same identifier are different between one repository and the other, as the metadata in the EBI repository undergoes curation. For example, the attribute names in NCBI records *lat_lon*, *geo_loc_name*, and *elev* are represented in EBI records as *latitude and longitude*, *geographic location*, and *elevation*, respectively. Certain attributes whose values in NCBI BioSample are, for example, *missing*, *N/A*, *null*, or variants, are completely absent in the EBI BioSamples counterpart records. The remaining attribute values seem to be unaltered. Overall, we found 17,680 user-defined attribute names in the BioSamples dataset.

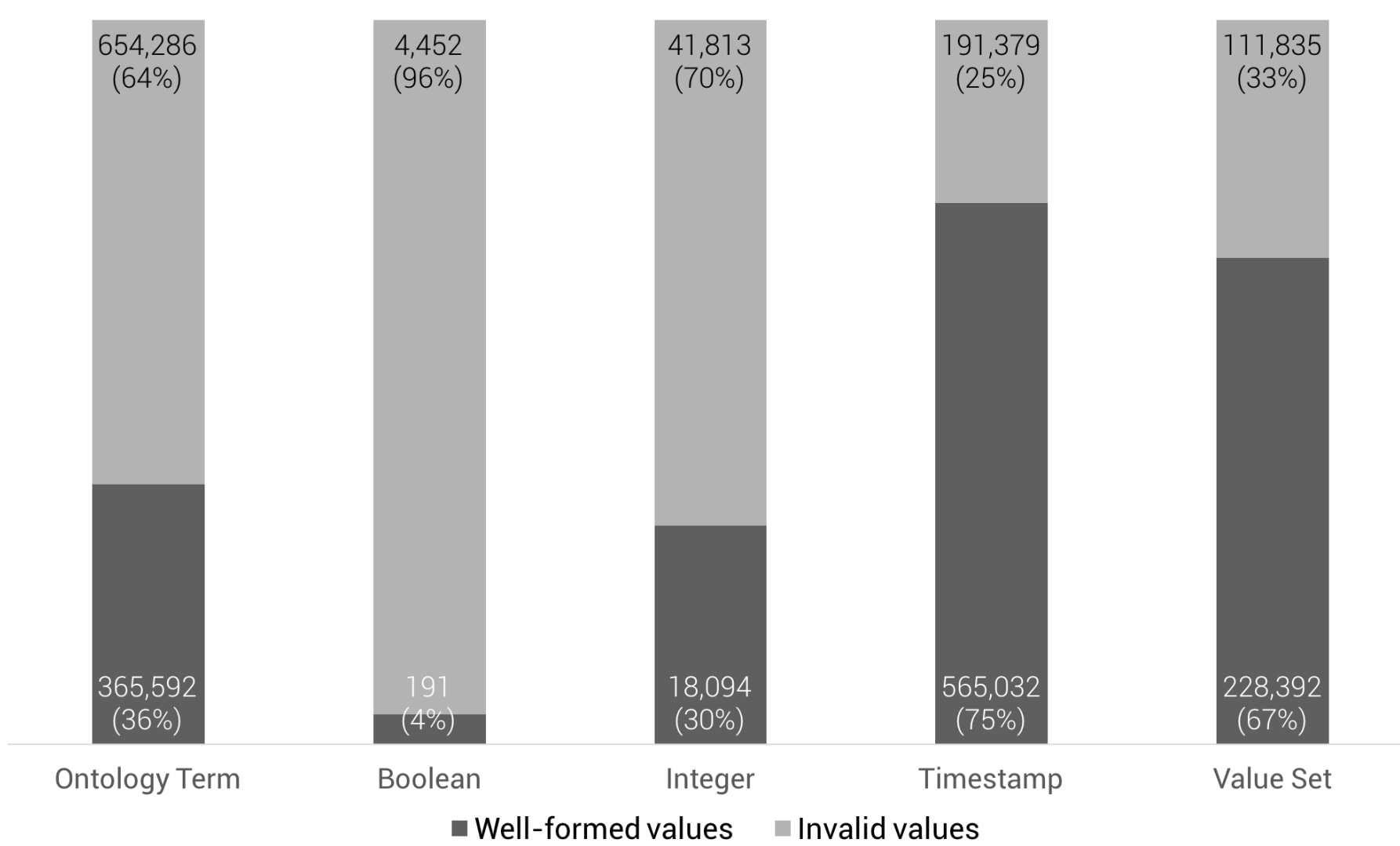

**Figure 5. Quality of attributes in packaged metadata records in NCBI BioSample**. The columns represent the metadata attribute types. Each column shows the number and percentage of metadata attributes whose values are either well-specified or invalid.

We investigated the quality of the metadata records occurring in both the NCBI and EBI databases. For the purposes of this investigation, we extracted a fragment of the NCBI

BioSample composed of the metadata records with identifiers that exist in EBI. We use the same quality criteria as for the NCBI repository, defined in Section 2.1. In **Figure 6** we show the results of our study. Observe that none of the attribute types is of superior quality compared to the results presented in Section 3.1 for the NCBI BioSample. The attributes that require a value from a value set have comparably significantly more invalid values.

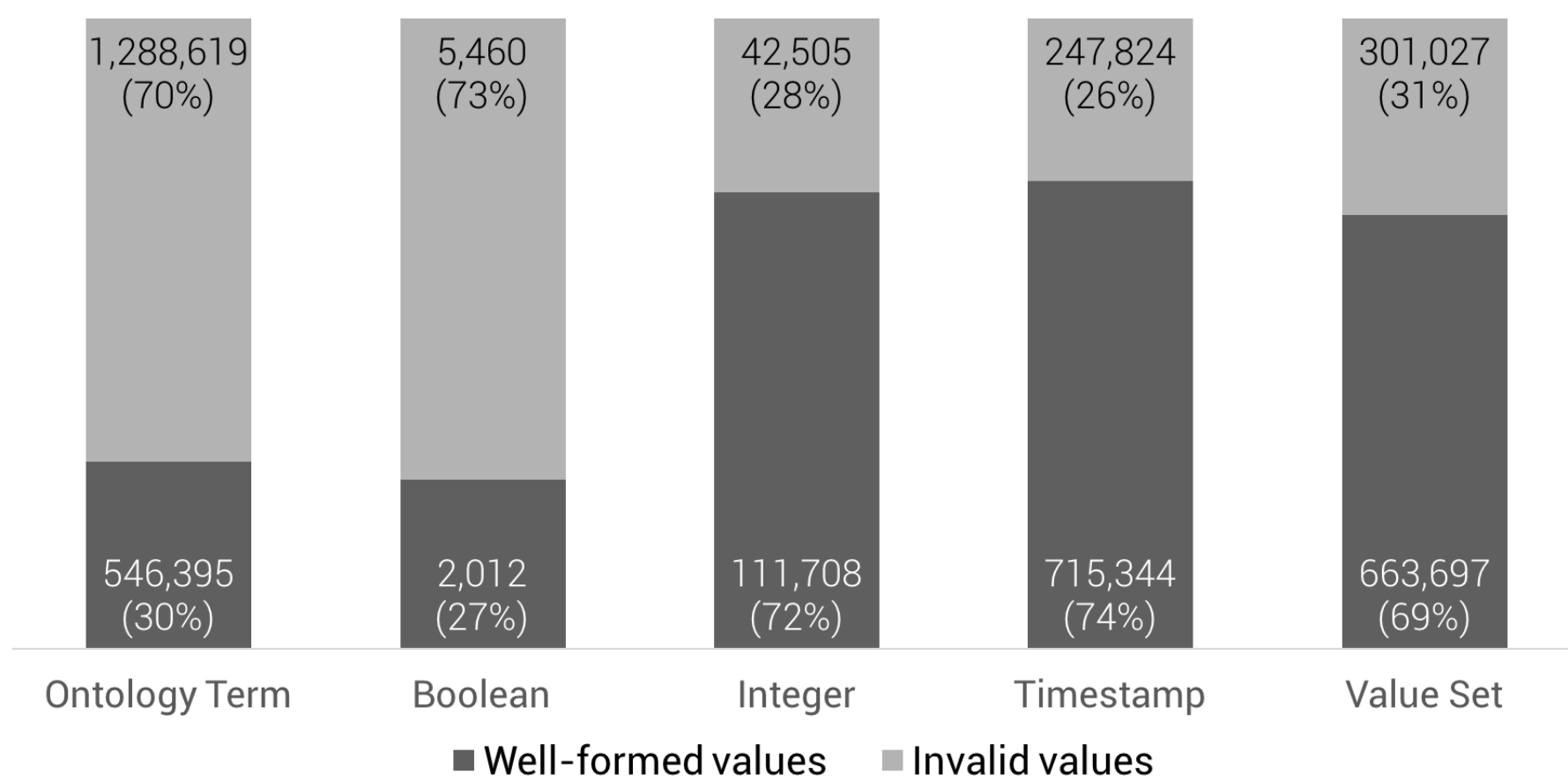

**Figure 6. Quality of attributes in metadata that co-exist in EBI and NCBI repositories.** The columns represent the metadata attribute types. Each column shows the number and percentage of metadata attributes whose values are either well-specified or invalid.

## EBI BioSamples

The EBI BioSamples repository overall contains a total of 4,793,915 metadata records. In **Figure 7**, we show the number of metadata submissions to the EBI BioSamples repository per year, from 2009 through to 2016.

The EBI BioSamples documentation does not reference the use of “package” specifications in the same way that NCBI BioSample does. However, we observed that 40% of the BioSamples metadata records (nearly 1.9M) contain package references; that is, they contain an attribute called *package* with a value that mirrors (or corresponds to) an NCBI package definition. The metadata records with a *package* attribute are potentially copied over from the NCBI BioSample. In **Figure 8**, we show the distribution of EBI BioSamples records according to whether they are unpackaged, or purport to adhere to a particular package. A large proportion of the packaged records adhere to the Generic package (41% of packaged records, 16% of total records). The next most used package is Metagenome/environmental (10% of packaged records, 4% of total records).

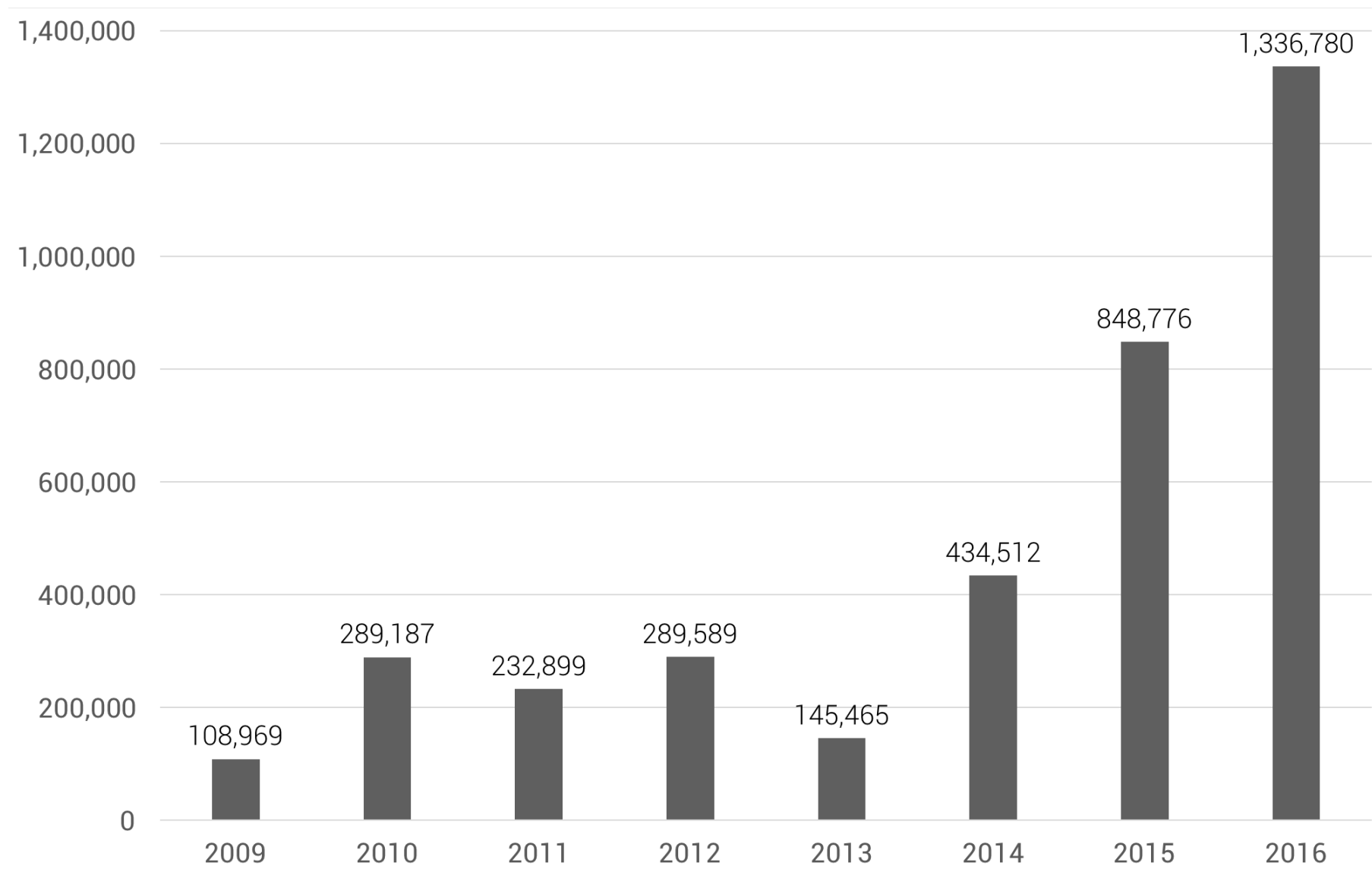

**Figure 7. Metadata submissions to EBI BioSamples from 2009-2017.** The columns represent the total number of metadata record submissions to EBI BioSamples per year.

The nearly 5M EBI BioSamples metadata records specify a total of 50,075,425 attribute name–value pairs. On average, each metadata record contains 10 attributes to describe the sample. A total of 5,708,592 attributes (11% of all attributes) in BioSamples records use named attributes. The remaining 44M attributes (89% of all attributes) use attribute names specified by metadata submitters rather than by EBI. We identified a total of 29,751 syntactically unique custom attribute names used in BioSamples metadata. By analyzing the 3 attribute names specified by the EBI repository, we found that nearly 99% (4,731,341 records) of BioSamples metadata records contain one entry for the *Organism* attribute. In contrast, only 1% of the metadata records contain a value for the *Material* attribute, and 19% contain a value for the *Sex* attribute.

In **Figure 9** we show the main results of our study of EBI BioSamples metadata, examining the 3 named attributes specified by the developers of the repository.

**Organism.** The majority of values for the *Organism* attribute are well-specified. There are, however, 618,925 values (13%) for which an exact term search on BioPortal yielded no results. Upon closer inspection, the values for the *Organism* attribute stored in the BioSamples records reference 8 unique URIs for the corresponding ontologies. These 8 URIs indicate 3 ontologies: the NCBI Taxonomy, the Mosquito Insecticide Resistance Ontology (MIRO), and the New Taxonomy (NEWT) ontology[25] of the SWISS-PROT group (now Uni-Prot). The URIs for MIRO and NEWT, as found in the metadata, could not be resolved. However, the MIRO ontology is part of the OBO Library[26], and it is hosted in both BioPortal and OLS. There are 1,826 attributes that mention the MIRO ontology with a link to a non-existing file in the SourceForge version-control repository that the OBO Library used to use (before moving to GitHub). While seemingly no longer in use, the NEWT ontology is mentioned in 132 metadata attributes. Furthermore, there are 17 URIs that link to the OLS page for the NCBI Taxonomy, 72 URIs that link to the BioPortal page for the NCBI Taxonomy, and 47 invalid URIs that are meant to link to the NCBI Taxonomy but do not have a colon following "http".

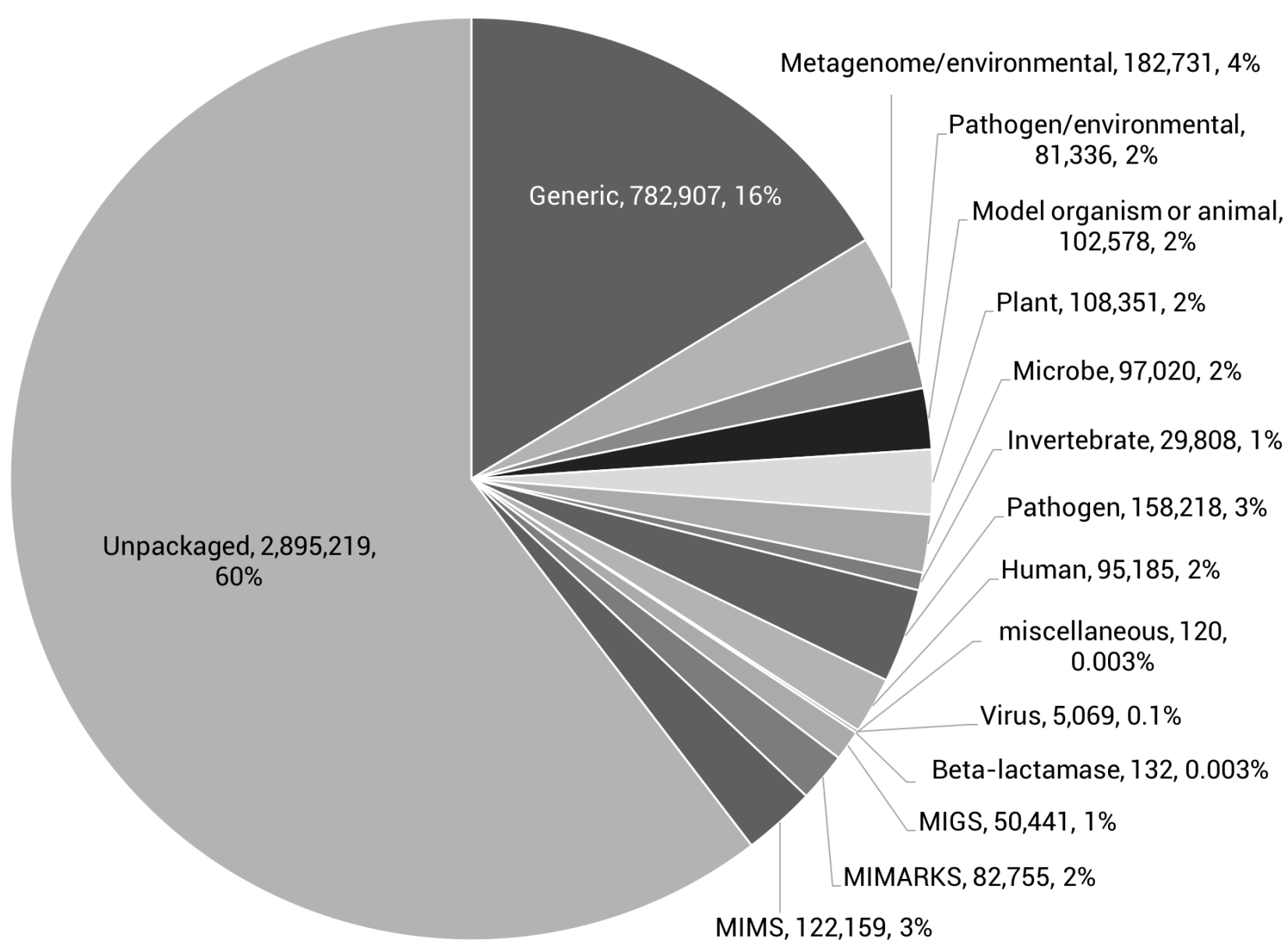

**Figure 8. Mention of metadata packages in EBI BioSamples.** The chart shows the package names (or "Unpackaged" for records that do not specify a package) followed by the number and percentage of metadata records that specify that package name.

**Material.** All but a few values for the *Material* attribute are well-formed. There are only 13 unique values for the *Material* attribute, 2 of which (stated in 4 metadata records) could not be found in BioPortal: *Mammillaria carnea rhizosphere*, and *primary tumour*. We did not get any results from an exact search for these values in the OLS ontology term search.

**Sex.** Most of the values for the *Sex* attribute are well-formed. Only 10% of the values did not resolve to any ontology terms in BioPortal. The invalid values we discovered include variations of the preferred "male" and "female", such as *male (XY)*, *males*, and *female (fertile)*. Other values that did not align with ontology terms include *mating_type_a,*

*unknown_sex, mixed_sex, gilt, 5 months, MATalpha, w, h-, u, B, F Age: 63, V, M Age: 69, XX, 77/M*, and multiple numbers and sentences.

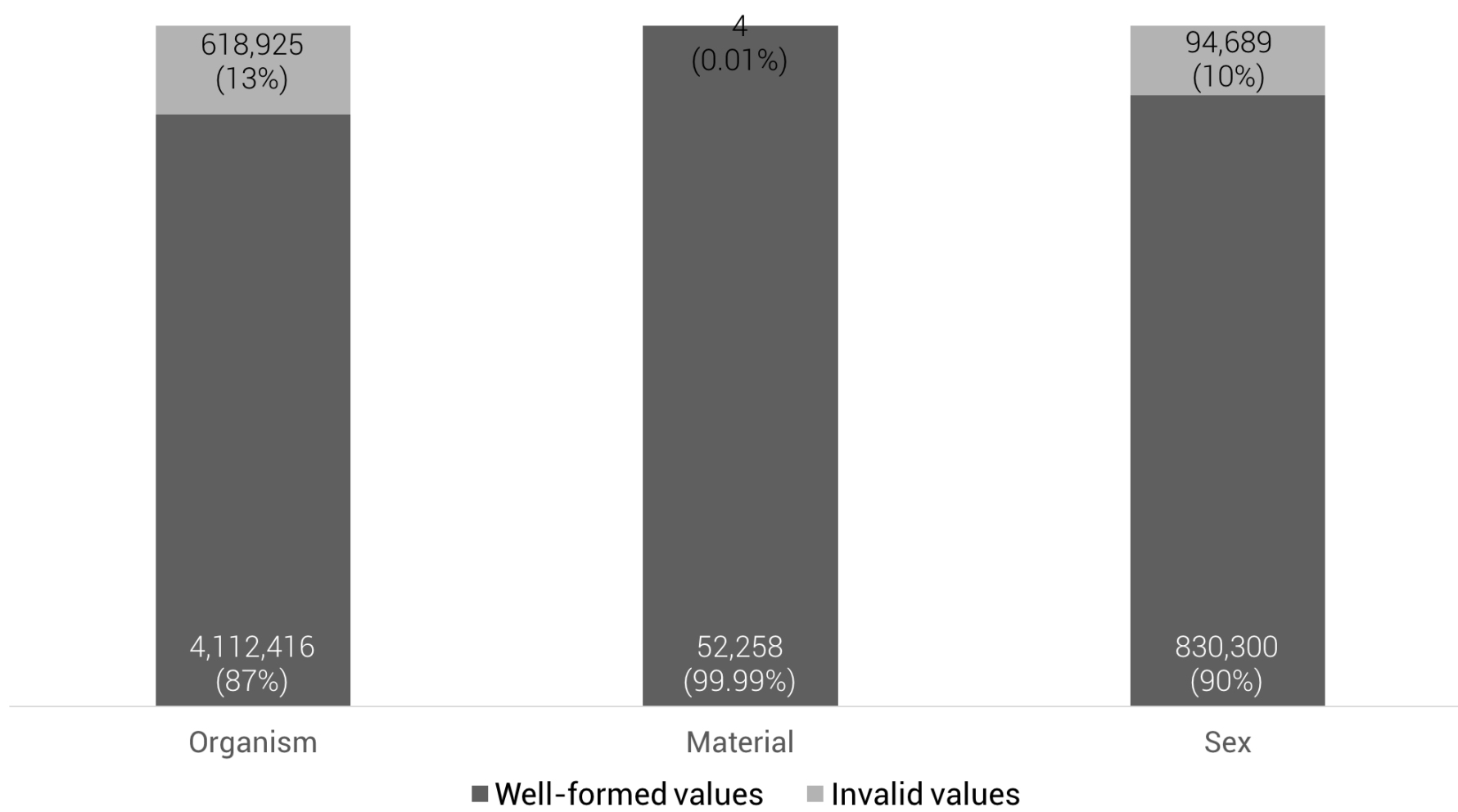

**Figure 9. Quality of named attributes in EBI BioSamples**. The columns represent the metadata attribute types. Each column shows the number and percentage of metadata attributes whose values are either well-specified or invalid.

The metadata values for named attributes seem to be generally very well-aligned with ontology terms. We found 82 syntactically distinct ontology URIs provided in metadata values for the three named attributes. Out of those, only 26 URIs can be resolved to an OWL[27,28] or OBO format[29] ontology. The 26 URIs resolve to 14 unique ontologies. For example, there were 6 different URIs for the EFO ontology—URIs using "http" or "https", ending with or without filename, ending with or without a slash). Out of the 14 resolvable URIs, 3 of them are links to specific ontology versions in a version-control system (GitHub and SourceForge) or FTP server.

## Clustering of metadata attribute names

We discovered in both the NCBI and EBI repositories a total of 33,143 syntactically unique attribute names. Out of those attribute names, there are 15,261 names that appear in metadata records of both the NCBI and EBI repositories. Among all these attribute names, we found by manual inspection that there were multiple attribute names used to represent the same aspects of a sample. For example, to represent the weight (in kilograms) of a sample, metadata submitters fabricate such attribute names as *weight (kg), weight_kg*, and *Weight..kg*. If all attributes denoting weight (in kilograms) used the same attribute name, software systems could rely on that single attribute name to precisely answer weight-related queries. This querying ability is currently impaired by the multitude of ways in which metadata submitters represent attributes of samples.

We therefore set out to find clusters of attribute names based on their similarity, which we compute using Levenshtein edit distance as our distance metric between attribute names. The Levenshtein distance is a standard metric used for spelling correction, and it is generally used in applications that benefit from soft matching of words (such as search systems). Our goal is to find clusters of terms that are used to represent the same aspect of a sample, by attending to their edit distance. For this purpose, we surveyed clustering approaches that (a) find exemplar values for each cluster, and (b) do not require an upfront specification of the number of clusters. These criteria rule out popular clustering methods such as *k*-means or *k*-medoids. We found that affinity propagation was the most applicable clustering algorithm for our analysis, satisfying both desiderata. By running our similarity detection tool based on affinity propagation, we identified 2,279 clusters of NCBI attribute names and 3,936 clusters of EBI attribute names.

In **Table 1** we show examples of the clusters produced by applying our tool to metadata attribute names.

**Table 1. Examples of clusters of metadata attribute names.** The left column contains the exemplar attribute name computed by the clustering algorithm, followed by the cluster of attribute names formed around the exemplar in the right column.

| Exemplar | Attribute names in cluster |
|---|---|
| atm pressure | atm press, atmospheric pressure |
| Disease stage | disease_stage, disease_stage, DiseaseStaging, disease staging, tfc_disease_stage, disease/status, diease_stat, DiseaseLocati |
| embryonic stage | embryo age, embryo stage, embryogenesis stage, embryonic age, embryonic day, embryonic stages, embryonic zone, meiotic stage, pollen embryo sac stage |
| environmental history | EnvironmentalHistory, Host Environmental History, environemental history, environmental history colony |
| experimental condition | Experiment condition, Experimental or control, enviromental conditions, environmental condition, environmental conditions, experimental conditions |
| genetic background | cytogenetic background, genetic and mutant background, genetic background cultivar, genetic backround, genetick background, genotype variation background |
| genotype variation | genotype variaion, cmv genotype variation, geneotype variation, |
| | genome variation, genoptype variation, genotype varation, genotype variaion, genotype variarion, genotype variataion, genotype variatation, genotype variaton, gentotype variation |
| geo_loc_name | geo_lac_name, geo_loc_name2, geo_loc_name_coord, geo_log_name, geooc_name, go_loc_name |
| nucleic acid extraction | Nucleic acid preparation, Nucleic_acid_extraction, nucleic acid amplification, nucleic acid extraction method |
| Sampling days | Sample Time days, Sampling Time  days approx, Sampling Year, Sampling days, sampling day |
| Submitted by | Submitter, Submitters |
| Time point | Time local, Time weeks, TimePointC, TimePointF, time point, time points, timepoint, timepoints |

In the first cluster for *nucleic acid extraction,* there is at least one attribute name that could clearly be used interchangeably with the exemplar attribute name *Nucleic_acid_extraction*. The remaining attribute names, while related to the exemplar attribute name, describe different aspects of the sample such as preparation and amplification of nucleic acids.

We computed the frequency of use of all attribute names in both the NCBI and EBI repositories. We then categorized the top 50 most widely used attribute names according to the concept that they represent. We chose to analyze only 50 attribute names as a way to sample meaningfully from the high number of attribute names. The resulting categorization, which is made up of disjoint categories, shows the most typical kinds of information conveyed in metadata records. In **Table 2** we show our main categories and examples of the attribute names that we grouped under each category.

**Table 2. Categories of attribute names according to the concept they represent.** The table shows the category in the left column, and the attribute names in that category in the right column.

| **Category** | **Attribute names** |
|---|---|
| Biomedical characteristic | breed, ethnicity, host, sample_type, organism, tissue, species, strain, sex, body site, cell type, genotype, disease state, … |
| Date | collection date, collection timestamp, time point |
| Geographic location | geo_loc_name, geographic location, lat_lon, country, latitude and longitude, grographic location (country and/or sea) |
| Measurement | depth, elevation, age, altitude, host_age |
| Identifier | sample id, package, model, gap_accession, gap_sample_id, … |
| Textual description | Sample_title, project name, Sample Name, label, title, study name, common name, secondary description, source name, … |

In the following analysis, we picked at least one attribute name from each of the categories in **Table 2**, and we determined how many similar attribute names exist in the metadata that could be used interchangeably with the selected attribute name. We did this by manually analyzing the clusters of attribute names, and identifying groups and subsets of clusters that, while their elements are syntactically different, can be seemingly used to represent the same aspect of a sample. The results of our analysis of attribute names based on their clusters is presented in **Table 3**.

**Table 3. Groups of attribute names seemingly used to describe the same concept.** From left to right, the table shows in each row: the concept that the metadata attributes presumably represent, the number of attribute names found to represent that concept, example attribute names found using our clustering method, and the numbers of metadata records in the NCBI BioSample and EBI BioSamples that contain attributes using one of the attribute names in the cluster. The standard attribute names specified in the NCBI BioSample documentation are shown in bold.

| Concept | #Attribute names | Example attribute names | #NCBI records | #EBI records |
|---|---|---|---|---|
| Geographic location | 32 | **latitude and longtitude**, lat_lon, Lat-Long, geo_loc, lat_lan, lat_lo, lat_lon (N:W), lat_lon_2, lat_long, latitude_longitude, geographic location (latitude and longitude), geographic location, geographic_location | 1,056,519 | 442,950 |
| Height | 31 | **height or length**, height cm, Height..cm., height, Height, height (cm), height in, Height (m), height m, height_cm, height_m, height_meters, Height in Centimeters, height_meter, height_meters | 23,170 | 23,641 |
| Elevation | 13 | **elevation**, elevation(m), Elevation m, elevation meters, gps elevation, geographic location (altitude/elevation), geographic location (elevation), estimated elevation | 119,477 | 157,778 |
| Age | 33 | **age**, age in years, age (in years), Age(years), age_in_years, age years, age (years), age(years), age_years, age (yr), age (yrs), age | 553,523 | 711,747 |
| | | in days, age_days, Age(days), AgeDays, age (days), age in weeks, age (weeks) | | |
| Weight | 26 | **total_mass**, weight kg, Weight..kg., weight (kg), weight_kg, weight, Weight_lb, weight lbs, weight_pounds, weight_g, Weight (g), weight_in_grams, Weight (mg) | 16,330 | 11,966 |
| Birth date | 18 | **birth_date**, date of birth, Date of birth, date_of_birth, year of birth, year_of_birth, birthyear, birth_year, year born, birth year | 22,785 | 19,684 |
| Time point | 62 | Timepoint, Time.point, time point, time points, time-point, time_point, Timepoints, time-point in minutes, timepoint in minutes, time-window, time, time_period, time period, time_point_days, time_point_months | 76,561 | 105,083 |
| Country or Region | 24 | **geographic location**, country region, CountryOrRegion, country location, country nation, country of origin, geographic location (country and/or sea), geographic location (country and/or sea, region), geographic location country 2 | 190,718 | 201,655 |
| Collection date/time | 32 | **collection_date**, collectiontime, collected date, collection_date (dmy), collection day, collection month, collection date unformat, collection time, collection timestamp | 136,819 | 139,231 |
| Ethnicity | 4 | **ethinicity**, ethinity, raceethnicity, ethnicity | 41,007 | 73,997 |
| Sample type | 31 | **sample_type**, sample type, sample-type, type of sample, type_sample, sample_ type, type sample, type_of_sample, sample type description, sample_type_beta | 260,708 | 299,868 |

In **Table 3**, there exists a standard attribute name, denoted in bold, to represent the concept in each row. For example, out of all 33 different attribute names to represent *age*, there are 32 custom (user-defined) attribute names and 1 standard attribute name defined by BioSample, which is *age*. While BioSample does not specify a standard attribute name for a time point, it documents specific attribute names for particular time points such as the time of sample collection, which should be described using the *collection_date* attribute.

## Discussion

We carried out an empirical assessment of the quality of metadata in two well-known online repositories of metadata about samples used in biomedical experiments: the NCBI BioSample and the EBI BioSamples.

Our study of the NCBI BioSample repository revealed multiple, significant anomalies in the metadata records. While NCBI BioSample promotes the use of specialized packages to provide some control over metadata submissions, the vast majority of submitters prefer to use the Generic package, which has no controls or requirements. A significant proportion of the attributes (15%) in NCBI's BioSample records use *ad hoc* attribute names that do not exist in BioSample's attribute dictionary. These 18,198 custom attribute names account for the clear majority of the attribute names (97.6%) used in metadata records, signaling a need to standardize many more than the 452 attribute names specified by BioSample. A considerable number of ontology-term attributes (68%) have values that do not correspond to actual ontology terms. The Boolean-type attributes have a staggeringly wide range of values, with only 27% of them being valid according to the NCBI BioSample specifications.

In our study of the EBI BioSamples, we discovered metadata of significantly higher quality compared to NCBI BioSample. The curation that the EBI BioSamples metadata undergo seems to produce high-quality alignments between the raw sample metadata values and ontology terms, albeit only in specific attributes. Thus, with appropriate tooling and appropriate standards, metadata can be significantly improved, even after submission. However, we found in EBI BioSamples an even higher degree of heterogeneity regarding custom attribute names—a total of 29,751 user-provided attributes. The vast majority of attributes (89%, 44M) in EBI BioSamples metadata use custom attribute names. While the curation applied to a few specific EBI BioSamples attributes results in high-quality values (for those attributes), all other attributes are populated with values that go by unchecked.

Our results demonstrate that the use of controlled terms from standard ontologies in sample metadata is rather sporadic, especially in the NCBI BioSample—a relatively modern initiative that aims at encouraging the standardization of its metadata. This situation ends up hampering search and reusability of the associated datasets. Although the requirements for NCBI BioSample metadata are well-specified, these requirements do not seem to be enforced during metadata submission. The result is clear: We observed that the metadata in NCBI BioSample are generally non-standardized and potentially difficult to search, and that the underlying repository suffers from a lack of appropriate infrastructure to enforce metadata requirements. The use of ontologies is particularly substandard, and even simple fields that require Boolean or integer values are often populated with unparsable entries.

The attribute-name clustering methods that we used still require some human curation effort to improve and to verify the resulting clusters, although such methods are surprisingly helpful to assist a human in sifting through the high number of related attribute names in the metadata that we studied and to quickly identify clusters of attribute names that denote the same concept. We identified multiple clusters of highly used idiosyncratic attribute names that could be represented using a single standard name per cluster. For example, we found 33 ways to represent age, 31 ways to represent height, and 32 ways to represent geographic locations (via their latitude and longitude). The attributes that we analyzed in our clustering study are frequently used in the metadata, and so we expect that those attributes are commonly used in searches that scientists perform. Because it is impossible for a searcher to anticipate all the variants that metadata authors might use, standardization of attribute names is particularly important if our goal is to make online datasets FAIR.

Overall, the multiplicity of attribute names used to describe the same thing (even in the same database) is highly detrimental to the searchability of the metadata, and, consequently, to the discoverability of the data that the metadata describe. With 32 different ways to represent the collection time of a sample, a scientist needs to cater to at least those many representations to find, for example, samples collected in the last year (if that information is provided in the metadata at all, though that is a separate problem). Finding data and metadata is and will continue to be problematic as long as metadata exhibits such a high degree of representational heterogeneity. Even though a scientist could identify the intention behind non-compliant data, finding that exact data is still a non-trivial challenge.

Our work suggests that there is a need for a more robust approach to authoring metadata. To be FAIR, metadata should be represented using a formal knowledge representation language, and they should use ontologies that follow the FAIR principles to standardize the metadata attributes and their values. These aspects help to ensure interoperability of the metadata, and are crucial for finding online datasets based on their metadata. The tooling available to scientists who author metadata should impose appropriate restrictions on the metadata. For example, wherever a value should be a term from a specific ontology, the metadata author should be presented only with options that are valid terms from that ontology when filling in metadata. We discussed the results of our work with the EBI BioSamples and the NCBI BioSample teams, in an effort to tighten the quality control of the metadata submitted to those repositories going forward. As for existing metadata, we are discussing with the EBI and the NCBI teams mechanisms to clean the metadata about biological samples in their repositories at scale.

Our findings guide the implementation of a software system that aims to complement the way scientists author metadata to ensure standardization, completeness, and consistency. The Center for Expanded Data Annotation and Retrieval (CEDAR)[30] is developing a suite of tools—the CEDAR Workbench[31–34]—that allows users to build metadata templates based on community standards, to fill in those templates with metadata values that are appropriately authenticated, to upload data and their metadata to online repositories, and to search for metadata and templates stored in the CEDAR repository. The goal of CEDAR is to improve significantly the quality of the metadata submitted to public repositories, and thus to make online scientific datasets more FAIR.

**Limitations and generalizability**

The investigation we carried out did not exhaustively evaluate the metadata in either the EBI or the NCBI repositories—we limited ourselves to select groups of attribute names that (a) are specified and documented by the repository developers, and (b) are computationally verifiable in an unambiguous manner. For example, to determine whether a complex value such as a measurement is valid, it is necessary to parse out the numeric value as well as the representation of the unit of measurement, which could be encoded in multiple ways (*Kg, kilograms, (in Kgs),* and so on).

In our clustering experiment, we used the Levenshtein edit distance—typically used in spell checkers—as our distance metric. This metric is agnostic to the semantics of terms, so using edit distance as a basis for clustering is unlikely to produce a clustering where, for example, *mass* and *weight* are in the same cluster—even though they represent the same attribute. We carried out a preliminary experiment where we computed the Euclidean distance between vector representations of metadata attribute names before clustering. However, the resulting clusters using affinity propagation were of worse quality than those computed using the Levenshtein edit distance.

To generalize our study to arbitrary metadata values, we would need to automate our methods to detect data types for user-defined fields, and to develop a mechanism to detect patterns in the metadata. Based on that information, we could automate the decision of whether a value is testable or not. For example, if the lengths of the values given for a single attribute are evenly distributed between 50 and 500 characters, then the field is plausibly a textual description, which is unlikely to have a correspondence with any controlled terms (though the value could certainly still be annotated with some ontology). On the other hand, if the values for an attribute exhibit date-like patterns, the attribute can be automatically verified using standard date-time formats.

**Acknowledgments**

This work is supported by grant U54 AI117925 awarded by the U.S. National Institute of Allergy and Infectious Diseases (NIAID) through funds provided by the trans-NIH Big Data to Knowledge (BD2K) initiative (http://www.bd2k.nih.gov). BioPortal has been supported by the NIH Common Fund under grant U54 HG004028.

We thank Marcos Martínez-Romero, Martin O'Connor, and John Graybeal (from CEDAR) for helpful discussions; Ben Busby and Tanya Barrett (from the NCBI) for useful exchanges; Helen Parkinson, Simon Jupp, and Melanie Courtot (from the EBI) for valuable discussions and exchanges, and for reviewing this manuscript.

**Author Contributions**

RSG: Study design, corpus collection, software implementation, study execution, data analysis, and manuscript writing.
MM: Conceptualization, study design, manuscript review and editing.

**Competing Interests**

The authors declare no competing interests.